# Detection and measurement in the V-band of the white dwarf spin period in the January 2004 outburst of DO (YY) Draconis

**David Boyd***

British Astronomical Association, Variable Star Section

A modulation in the V-band with period 527.84±1.81 sec and amplitude 0.023 magnitude, attributable to the spin of the magnetic white dwarf primary star, has been detected in 7.5 hours of V-band CCD photometry data recorded during the January 2004 outburst of the DQ Her type dwarf nova DO (YY) Draconis. This measurement is consistent with previous results for the white dwarf spin period based on X-ray and UV observations made with the Hubble Space Telescope, ROSAT and the Rossi X-ray Timing Explorer and appears to be the first independent determination of the spin period in the V-band. It is consistent with previous V-band observations in not showing a significant signal at the first harmonic of the spin period as seen in X-ray and UV data. Light output in the V-band peaks only once per rotation of the white dwarf rather than twice as seen in X-rays and UV. Concurrent observation at optical and X-ray wavelengths is needed to establish whether these two modulation behaviours occur at the same time during the outburst and to investigate the phase relationship between them. A coordinated observing campaign at a future outburst would help to advance our understanding of this system.

## Nature of DO Dra

DO Dra is classified in Downes[1] as a DQ Her intermediate polar type of cataclysmic variable (CV). In these close binary systems, the white dwarf primary star has a sufficiently strong magnetic field that it disrupts the inner portion of the accretion disc. Rather than the material in the disc gradually spiralling down onto the primary as in normal CVs, the magnetic field causes the material at the inner edge of the disc to flow in two streams onto the two magnetic poles of the white dwarf. This material is accelerated and impacts the poles at around 3000 km/sec, emitting strongly in X-rays and UV light as well as at optical wavelengths.

In the case of DO Dra, these magnetic poles are believed to lie close to the equatorial plane. As the white dwarf spins, the changing configuration of these light-emitting regions above the magnetic poles relative to our line of sight causes modulation of the light output from the system. By detecting this modulation, it is possible to measure the white dwarf spin period. For a fuller description of these systems see, for example, Hellier.[2] The orbital period and orbital inclination of DO Dra are reported in Ritter & Kolb[3] as 0.165374 days (3.97 hours) and 45° respectively. Its distance is reported in Mateo[4] as 155 parsec.

## Identity of DO Dra

There has been some confusion about the identity of the variable catalogued as DO Draconis. An apparent eclipsing binary discovered near this position by Tsesevich in 1934 and subsequently catalogued as YY Dra is no longer visible, either due to incorrectly recorded coordinates or to misidentification of its nature. The CV currently observed at this location has now been catalogued in Downes[1] and Ritter & Kolb[3] as DO Dra. For a discussion of this issue, see Patterson.[5] In this paper, the variable will be referred to as DO Dra.

## Previous observations

Previous observations of periodicity in DO Dra at X-ray, UV and optical wavelengths are listed in Table 1, together with the reference number in this paper of the relevant publication. In each case the period listed is for the strongest periodic signal detected.

The $CuSO_4$ filter is essentially a red-blocking filter and transmits all wavelengths below about 700nm.

**Table 1. Previous observations of periodicity in DO Dra**

| Date | Instrument | Wavelength | Period | Ref. |
| --- | --- | --- | --- | --- |
| 1990 Jan | KPNO 0.9m | ($CuSO_4$) | 265.8±1.3 sec | 5 |
| 1991 Nov | ROSAT PSPC | X | 265.1±1.8 sec | 6 |
| 1992 Oct | ROSAT HRI | X | 264.1±1.5 sec | 6 |
| 1994 Jun | HST FOS | UV | 264.71±0.05 sec | 7 |
| 1994 Jun | McDonald 2.1m | U | 264.6±0.8s | 7 |
| 1996 May | ROSAT HRI | X | 264.7±0.1 sec | 8 |
| 1999 Sep | RXTE | X | 264.66 sec * | 9 |
| 1999 Sep | Washington 0.76m | V | 529.31 sec * | 9 |
| 2000 Nov | RXTE | X | 264.66 sec * | 9 |
| 2000 Nov | Sternberg 0.6m | V | 529.31 sec * | 9 |
| 2000 Nov | Crimean 0.38m | V | 529.31 sec * | 9 |
| 2000 Nov | CBA Concord 0.44m | Unfiltered | 529.31 sec * | 9 |

The 1999 and 2000 observations reported in Szkody[9] and marked with * used the period listed in the table, which was published as a weighted mean of previous observations in Haswell,[7] to analyse their data but did not publish a value derived from their own data.

The interpretation generally adopted is that the signal at 265 sec seen strongly in X-rays and UV is the first harmonic of a white dwarf sidereal spin period of 530 sec. The detection of two peaks per rotation of the white dwarf is attributed to alternate visibility of the accretion regions above the two magnetic poles. There is a discussion in Norton[8] on how the strength of the magnetic field of the white dwarf might affect the geometry of the infalling material above the poles and thereby determine whether one or two peaks are visible per rotation. The conclusion drawn there is that visibility of two peaks may indicate that DO Dra has a relatively weak magnetic field.

It is interesting to note that the V-band observations in 1999 and 2000 only detected a signal consistent with a period of 529 sec, although the modulation seen was a marginal detection. No signal in the V-band was detected at 265 sec.

## New observations

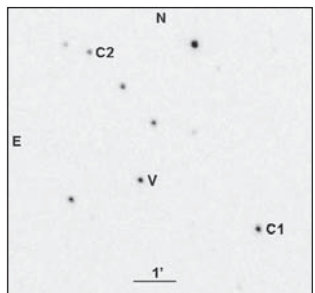

**Figure 1.** Field of DO Dra.

DO Dra was reported to be in outburst at magnitude 14.5 by Mike Simonsen on 2004 January 23.[10] On the following evening, 24/25 January, a 7.5 hour sequence of CCD images of the DO Dra field was recorded using a 25cm telescope, HX516 CCD camera and V-band filter. Figure 1 shows the field during outburst with the variable (V), comparison (C1) and check (C2) stars marked.

Astrometry of DO Dra using *Astrometrica*[11] with the USNO B1.0 catalogue provided the position RA 11h 43m 38.51±0.16s, Dec +71d 41m 20.73±0.16s. All images were bias- and dark-subtracted and flat-fielded before being analysed photometrically using *AIP4WIN*.[12] This produced the lightcurve shown in Figure 2.

Star C1 (V 10.78, B–V 0.56) on AAVSO chart 1137+72(f)[13] was used as the comparison. The accuracy of individual

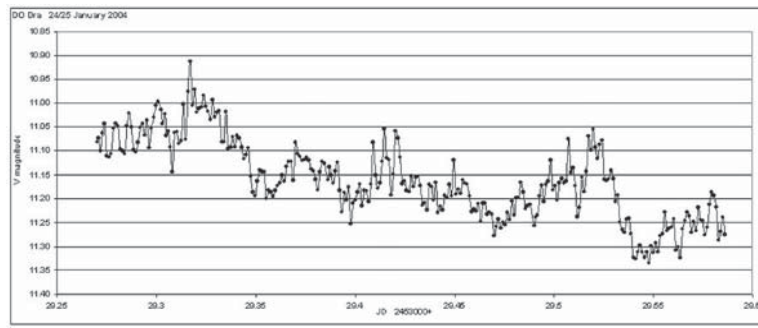

**Figure 2.** DO Dra V-band lightcurve covering 7.5 hr, 2004 January 24/25.

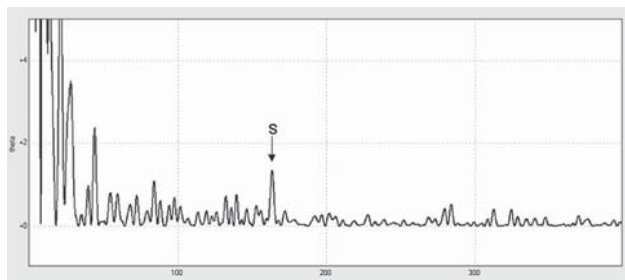

**Figure 3.** Lomb–Scargle power spectrum of DO Dra lightcurve.

DO Dra differential V magnitude measurements was computed as ±0.008 mag using the error formula in Howell.[14] This includes the effects of photon statistics, readout noise and dark current in the CCD. Measurements of the check star C2 (V 12.48, B–V 0.44) throughout the run were used to verify these error estimates. The average computed error on the check star over the run using the Howell formula was ±0.017 mag, while the standard deviation of check star differential magnitude measurements over the run was ±0.016 mag, in good agreement. The error bars on individual V magnitude measurements of DO Dra are too small to reproduce clearly in Figure 2 and so are omitted.

Figure 2 shows a high level of activity in the V-band. Although the peak to peak variation of these fluctuations is about 0.05 magnitude, and therefore considerably larger than the estimated error on an individual measurement, a further check that this variation is real and not due to unknown random errors was made by reanalysing all the images using a different comparison star, in this case star 4 on the AAVSO chart. The two lightcurves showed very similar behaviour with the difference between them having a standard deviation of ±0.0114 magnitude compared to an expected value of ±0.0111 based on the known errors of the individual magnitude measurements. The light curves of DO Dra produced using two different comparison stars were therefore essentially the same indicating that these fluctuations are indeed real and due to intrinsic variation in the light output of the DO Dra system. Similar flickering was observed during previous outbursts as noted in Szkody.[9]

## Data analysis

A Lomb–Scargle spectral analysis of these data using *Peranso*, a new period analysis package developed by Vanmunster,[15] gave the results shown in Figure 3. The peak in the power spectrum indicated with the label S is at a frequency of 164 cycles/day corresponding to a period of 528 sec.

As noted in other observations of DO Dra, the spectrum is dominated by low frequency variation of the lightcurve on which a relatively small high frequency variation is superimposed. To remove the low frequency component, a running average of length 0.02 days was calculated and subtracted from the measured lightcurve. The running aver-age lightcurve containing the low frequency variation is shown in Figure 4 along with an indication of the previously determined orbital period of 0.165374 days. No significant periodic signal was seen in the power spectrum at the corresponding frequency of 6.05 cycles/day.

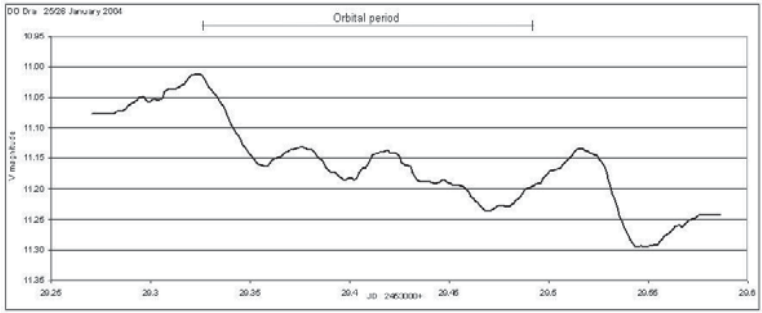

**Figure 4**. DO Dra running average lightcurve.

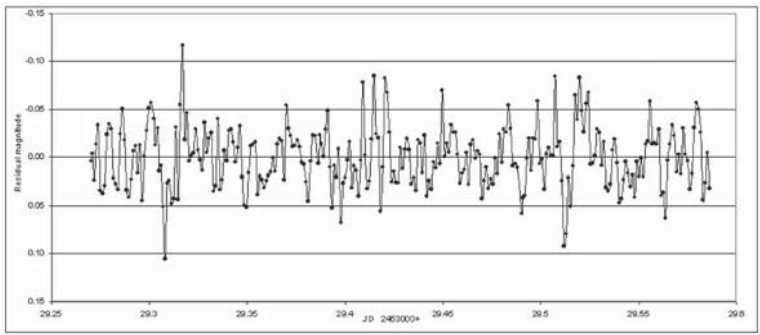

**Figure 5.** DO Dra residual lightcurve after subtracting running average.

The residual lightcurve containing the higher frequency variation is shown in Figure 5.

A Lomb–Scargle analysis of this residual lightcurve using *Peranso* gave the results in Figure 6. The previously detected peak is now the strongest signal in the data. Detailed analysis of this spectrum gave the frequency of the peak as 163.686±0.560 cycles/day corresponding to a period of 527.84±1.81 sec.

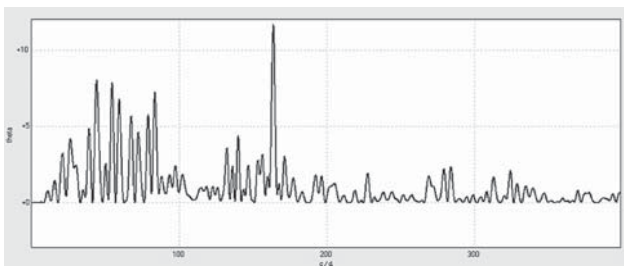

**Figure 6.** Lomb–Scargle spectrum of residual lightcurve after removal of low frequency variation.

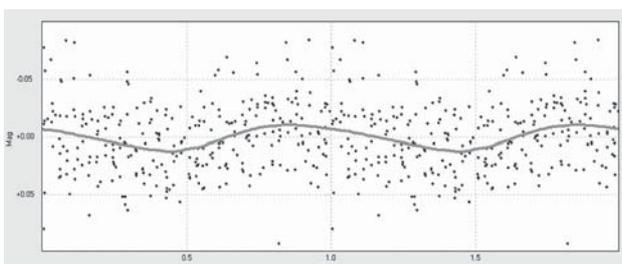

**Figure 7.** Data folded on measured white dwarf spin period.

As a further check, the residual lightcurve (Figure 5) was reanalysed using the Fourier transform routine of the *Period98* software.[16] This gave a value of 527.83 sec for the peak in excellent agreement with *Peranso*.

Using *Peranso*, the residual lightcurve was folded at the measured period of 527.84 sec. The result is shown in Figure 7 with a curve drawn through the mean values. This shows two cycles for clarity. There is clear evidence of modulation of the lightcurve at this period with an amplitude of 0.023 magnitude.

To check this result, the folded data were aggregated in bins of width 0.1 in phase and the mean values and standard deviations of each bin plotted in Figure 8. Again two cycles are shown. This confirms the modulation and amplitude reported by *Peranso*. As a further check on the robustness of this modulation, Figure 8 also shows the same data rebinned with a phase shift of 0.05.

## Comparison with previous observations

The measured values of the white dwarf sidereal spin period derived from the observations listed in Table 1, together with this measurement, are listed in Table 2 and plotted in Figure 9.

The value for the white dwarf sidereal spin period measured here, 527.84±1.81 sec, is consistent with previous measurements. While these results are entirely consistent with a constant spin period, as has been assumed in the past, there is also a hint that the period could be decreasing. Further observations with the accuracy achieved by HST or ROSAT would clarify this.

The modulation in the V-band measured here is more clearly defined than that shown in Figures 5 and 6 of Szkody.[9]

There is no evidence in the power spectrum of a significant signal at or near 265 sec (326 cycles/day), the first harmonic of the white dwarf spin period. This is consistent with earlier V-band observations which appear to show only one peak per rotation in contrast to X-ray and UV results which show two peaks.

**Table 2. Measured white dwarf spin periods for DO Dra**

| Date | Period | Wavelength | Ref. |
| --- | --- | --- | --- |
| 1990 Jan | 531.6±2.6 sec | (CuSO$_4$) | 5 |
| 1991 Nov | 530.2±3.6 sec | X | 5 |
| 1992 Oct | 528.2±3.0 sec | X | 6 |
| 1994 Jun | 529.42±0.1 sec | UV | 7 |
| 1994 Jun | 529.2±1.6s | U | 7 |
| 1996 May | 529.4±0.2 sec | X | 8 |
| 2004 Jan | 527.8±1.8 sec | V | This measurement |

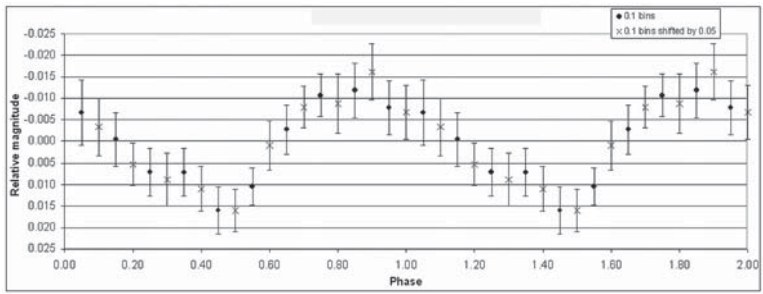

**Figure 8.** Folded lightcurve plotted in bins of 0.1 in phase.

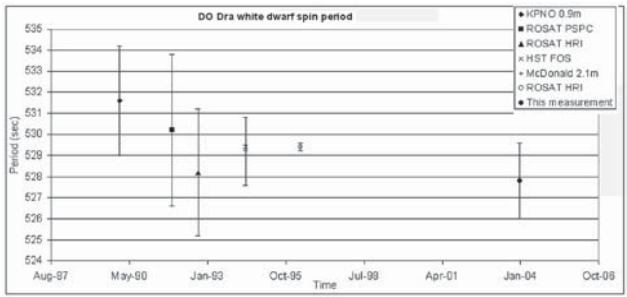

**Figure 9.** Measured white dwarf spin periods for DO Dra.

Norton[8] presents a possible explanation for single and double-peaked X-ray spin pulse profiles in terms of the strength of the magnetic field of the white dwarf. He argues that the presence of two peaks as for DO Dra is an indication of a relatively weak magnetic field. Hellier[2] comments that we currently lack sufficient observational evidence to understand why some intermediate polar systems show one peak per rotation and others show two. The situation is clearly more complex when the same system can apparently show both behaviours at different wavelengths.

# Conclusion

A modulation in the V-band with period 527.84±1.81 sec and amplitude 0.023 magnitude, attributable to the spin of the magnetic white dwarf primary star, has been detected during the January 2004 outburst of DO Dra. This appears to be the first independent determination of the spin period in the V-band and is consistent with previous observations using a variety of instruments including the HST, ROSAT and RXTE. It is also consistent with previous V-band observations in not showing a significant signal at the first harmonic of the spin period. Light output in the V-band peaks only once per rotation of the white dwarf rather than twice as seen in X-rays and UV. Concurrent observation at optical and X-ray wavelengths is needed to clarify whether single and double peaked modulation behaviour occurs at the same time during the outburst and to establish the phase relationship between them. In particular, it would be interesting to investigate the phase relationship between these observations and any X-ray or UV data obtained during the January 2004 outburst. Failing this, a coordinated observing campaign at a future outburst is needed.

# Acknowledgments

The author wishes to thank Tonny Vanmunster for providing a beta version of his *Peranso* period analysis software, Chris Lloyd at the Rutherford Appleton Laboratory for his encouragement and advice on analysing the data, both referees for their helpful comments and Ann Davies at the BAA Office for obtaining copies of some of the references.

**\*Address:** 5 Silver Lane, West Challow, Wantage, Oxon. OX12 9TX [drsboyd@compuserve.com]

# References

1 Downes R. A. *et al.*, *A Catalogue and Atlas of Cataclysmic Variables – Living Edition*, **http://icarus.stsci.edu/~downes/cvcat/** (2001)
2 Hellier C., *Cataclysmic Variable Stars: How and why they vary*, Springer−Verlag, 2001
3 Ritter H. & Kolb U., 'Catalogue of Cataclysmic Binaries, Low-Mass X-Ray Binaries and Related Objects', *Astronomy and Astrophysics*, **404**, 301, and RKcat Edition 7.2, **http://physics.open.ac.uk/RKcat/** (2004 January 1)
4 Mateo M. *et al.*, 'Near-infrared time-resolved spectroscopy of the cataclysmic variable YY Draconis', *Astrophysical Journal*, **370**, 370−383 (1991)
5 Patterson J. *et al.*, 'Rapid oscillations in cataclysmic variables. VIII. YY Draconis (=3A 1148+719)', *Astrophysical Journal*, **392**, 233−242 (1992)
6 Patterson J. & Szkody P., 'Rapid oscillations in cataclysmic variables. XI. X-ray pulses in YY Draconis', *Publications of the Astronomical Society of the Pacific*, **105**, 1116−1119 (1993)
7 Haswell C. A. *et al.*, 'Pulsations and accretion geometry in YY Draconis: A study based on Hubble Space Telescope observations', *Astrophysical Journal*, **476**, 847−864 (1997)
8 Norton A. J. *et al.*, 'YY Draconis and V709 Cassiopeiae: two intermediate polars with weak magnetic fields', *Astronomy and Astrophysics*, **347**, 203−211 (1999)
9 Szkody P. *et al.*, 'X-ray/optical studies of two outbursts of the intermediate polar YY (DO) Draconis', *Astronomical Journal*, **123**, 413−419 (2002)
10 Simonsen M., 'DO Dra in outburst', VSNET outburst notice 6123, 2004 January 23
11 Raab H., *Astrometrica*, **http://www.astrometrica.at/**
12 Berry R. & Burnell J., *The Handbook of Astronomical Image Processing*, Willmann−Bell, 2000
13 American Association of Variable Star Observers, **http://www.aavso.org/observing/charts/**
14 Howell S. B., *Handbook of CCD Astronomy*, p.53, Cambridge University Press, 2000
15 Vanmunster T., *Peranso*, **http://www.peranso.com**
16 Sperl M., *Period98*, **http://www.astro.univie.ac.at/~dsn/dsn/Period98/**